\documentclass[pra,twocolumn, nofootinbib, preprint,10pt,amsmath,amssymb, superscriptaddress,showkeys]{revtex4-2}
\usepackage[caption=false, subrefformat=parens, labelformat=parens]{subfig}
\usepackage{graphicx}
\usepackage[dvipsnames]{xcolor}
\usepackage{dcolumn}
\usepackage{bm}
\usepackage{amsmath, amsthm}
\usepackage{rotating}
\usepackage{amsmath,amsthm,amssymb,mathtools}
\usepackage{algorithm}
\usepackage[noend]{algpseudocode}
\usepackage{verbatim}
\usepackage{float}
\usepackage{outline}
\usepackage{tikz}
\usepackage{empheq}
\usetikzlibrary{shapes.geometric, arrows}
\usepackage{braket}
\usepackage{colonequals}
\usepackage{textcomp}
\usepackage[inline]{enumitem}
\usepackage[colorlinks]{hyperref}
\usepackage[all]{hypcap}
\usepackage[utf8]{inputenc}
\usepackage[english]{babel}
\usepackage{csquotes}
\usepackage{natbib}
\usepackage{caption}
\usepackage{soul}
\sethlcolor{yellow}
\usepackage[normalem]{ulem}
\usepackage[margin=0.7in]{geometry}

\begin{document}

\title{Mixer-Phaser Ansätze for Quantum Optimization with Hard
Constraints}
\author{Ryan LaRose}
\affiliation{Quantum AI Laboratory (QuAIL), NASA Ames Research Center}
\affiliation{USRA Research Institute for Advanced Computer Science (RIACS)}
\affiliation{Department of Computational Mathematics, Science, and Engineering, Michigan State University}

\author{Eleanor Rieffel}
\affiliation{Quantum AI Laboratory (QuAIL), NASA Ames Research Center}

\author{Davide Venturelli}
\email[]{davide.venturelli@nasa.gov}
\affiliation{Quantum AI Laboratory (QuAIL), NASA Ames Research Center}
\affiliation{USRA Research Institute for Advanced Computer Science (RIACS)}

\begin{abstract}
We introduce multiple parametrized circuit ansätze and present the results of a
numerical study comparing their performance with a standard Quantum Alternating
Operator Ansatz approach. The ans\"atze are inspired by mixing and phase separation in the QAOA, and also motivated by compilation considerations with the aim of running on near-term superconducting quantum processors. The methods are tested on random instances of a quadratic binary constrained optimization problem that is fully connected
for which the space of feasible solutions has constant Hamming weight. 
For the parameter setting strategies and evaluation metric used, the average performance achieved by the QAOA is effectively matched by the one obtained by a “mixer-phaser” ansatz that can be compiled in less than half-depth of standard QAOA on most superconducting qubit processors. 
\end{abstract}


\maketitle

\section{Introduction}\label{sec:intro}

The Quantum Approximate Optimization Algorithm (QAOA) was initially introduced in Ref.~\cite{farhi2014quantum}. Its simple structure inspired heuristic algorithms for sampling and exact optimization as well as approximate optimization that generalized the simple structure to include a broader and often more implementable set of operators.  

The algorithms following the ansatz alternate $p$ times between unitary operators chosen from a one-parameter family of \emph{phase separation} operators and operators chosen from a one-parameter family of \emph{mixing} operators. 
The mixing operators do not commute with the phase separation operators, enabling exploration of the search space. The aim is to output a state that has good overlap with the low-energy eigenspace of the problem Hamiltonian after $p$ layers. Good parameters can sometimes be determined analytically, or estimated efficiently classically, or may be found using a combination of runs on an quantum processing unit (QPU) together with a classical optimization heuristic~\cite{akshay2021parameter,rabinovich2021progress}. The simplest case, one which has been extensively studied in the QAOA literature in terms of theory, numerics, and experiments, is QAOA for $\mathtt{MaxCut}$, for which the ansatz alternates layers consisting of two-qubit parity gates $\mathcal{U}^{nm}_{ZZ}(\gamma)=\exp[i\gamma Z_nZ_m]$ with single qubit X-rotations $\mathcal{U}^{n}_{X}(\beta)=\exp[i\beta X_n]$ (\emph{X-mixer}s).

In~\cite{hadfield2019quantum}, the QAOA approach was generalized to the \emph{Quantum Alternating Operator Ansatz}, considering unitary layers that are not necessarily linked to local Hamiltonian evolution\footnote{In the remainder of this paper, we use the acronym QAOA to mean Quantum Alternating Operator Ansatz, which includes the Quantum Approximate Optimization Algorithms as a special case.}. In particular, multi-qubit mixing operators were introduced in lieu of the $X$-rotations when applying the QAOA to hard-constrained optimization in order to restrict the search to the feasible subspace, the space of valid configurations obeying the hard constraints.  
The simplest among the advanced mixers is the two-qubit \emph{XY gate}
\begin{equation}
\mathcal{U}^{nm}_{XY}(\beta)=\exp\left[i\beta(X_n X_m +Y_n Y_m)\right]
\end{equation} 
which conserves the total spin projection $Z_n + Z_m$. 

In this work, we introduce a new ansatz that combines the mixing and phase-separation operators into a more general two-parameter family of operators. For this reason, we refer to it as the 
"Quantum Alternate Mixer-Phaser Ansatz" (QAMPA).
The main motivation for this generalization is to reduce the depth of the circuits, potentially reducing performance in the ideal case (by potentially limiting the expressibility of the ansatz) but obtaining improved performance on noisy intermediate-scale quantum (NISQ) processors (by running shorter-depth circuits with that can still find good approximate solutions). We perform a numerical study on the performance of QAMPA
on a weighted combinatorial optimization problem with hard constraints. For fully-connected binary quadratic optimization problems, the circuits compile to roughly half the depth of standard QAOA on QPUs with nearest-neighbor connectivity. This is the case in most superconducting qubit quantum computers in which qubits are placed on a two-dimensional grid and interact with nearest neighbors through tunable or fixed frequency couplers. Our numerical simulations show that for the problems studied, in the noiseless case, QAMPA performs almost as well as standard QAOA in parameter regimes that are achievable in current hardware, and thus is expected to have advantages under noise given its reduced depth.
This ansatz is therefore a viable and attractive approach, particularly for highly connected optimization problems with hard constraints on NISQ hardware.

\section{Background and prior work}\label{sec:BG}

Experimental benchmarks with $X$-mixers are numerous, especially on superconducting processors, although mostly limited to problems whose topology exactly matches the quantum processor hardware (see~\cite{harrigan2021quantum} for a review). Ref.~\cite{harrigan2021quantum} also explores optimization of the fully connected Sherrington-Kirkpatrick model,which requires significant compilation overhead. Unsurprisingly, the compilation requirements resulted in significant performance degradation with circuit depth, due to the relentless unmitigated action of noise during circuit execution in NISQ hardware.

Many techniques have been developed to optimize gate synthesis and qubit routing (i.e. compilation) for algorithms to be run on noisy intermediate scale quantum (NISQ) devices featuring a sparse native gate set. Although experimental QAOA work with XY mixers has still not appeared in the literature, numerical analyses predict long circuit durations for problems with XY mixers that are not encouraging for very near-term hardware~\cite{do2020planning}. In \cite{farhi2017quantum, kandala2017hardware} hardware-efficient ans\"atze were proposed that match the processor topology and the native gates and use the objective function Hamiltonian only to guide the parameter setting procedure and to evaluate the final performance metric. In that approach, QAOA was used as a form of a quantum neural network that needs to be trained to act as an optimization solver, but there is concern as to how well this method, with its many parameters, would work in general.

The standard QAOA approach applied to combinatorial search was discussed in detail in \cite{hadfield2019quantum}. For a given cost function, it starts from a superposition ideally equally distributed among all possible candidate solutions, i.e.
\begin{equation}
    |\psi_0\rangle=|\mathcal{F}|^{-\frac{1}{2}} \sum_{s\in \mathcal{F}} |s\rangle
    \label{eq:init}
\end{equation}
where $\mathcal{F}$ represents the \emph{feasible subset} of the optimization problem.
This state is evolved to a quantum state $|\psi_F\rangle$ through a circuit composed by alternating sequentially two layers of gates for a number $p$ of rounds. Each round consists of an exploitation (phase-separation) layer $\mathcal{U}_{PS}(\gamma)$ which introduces information related to the cost function to be extremized and an exploration (mixing) layer $\mathcal{U}_{M}(\beta)$ which rearranges probability amplitudes across $\mathcal{F}$. The parameters $\gamma$ and $\beta$ are real numbers that need to be optimized layer-by-layer.
In practice these layers can be decomposed by products of single and two-qubit gates in an arbitrary order%
\footnote{Note that in general the order doesn't matter for phase separation, while $[\mathcal{U}_{M}^{nm}(\beta),\mathcal{U}_{M}^{kl}(\beta^\prime)]\neq0$, so ordering is important for mixing gates.}%
, e.g.  $\mathcal{U}_{PS}(\gamma)=\prod_{n, m}\mathcal{U}_{PS}^{nm}(\gamma)$.
A final quantum alternating operator ansatz looks like:
\begin{eqnarray}
    |\psi_F(\vec\gamma,\vec\beta)\rangle&=&
    |\mathcal{F}|^{-\frac{1}{2}}
    \prod_{i = 1}^{p} \left[\mathcal{U}(\gamma_i, \beta_i)\right]    \sum_{k\in\mathcal{F}}|k\rangle\\
    \mathcal{U}(\gamma_i,\beta_i)&=&
    \left(
    \prod_{n,m}\mathcal{U}_{M}^{nm}(\beta_i)
    \prod_{n,m}\mathcal{U}_{PS}^{nm}(\gamma_i)
    \right).
    \label{eq:ansatz}
\end{eqnarray}

Note that these products must be ordered when the two-qubit gates do not commute. Most of the works on QAOA feature exclusively single qubit gates as mixers. 
Only a few works have discussed the performance of the QAOA using $XY$-mixers: Ref.~\cite{wang2019xy} studies $\mathtt{MaxkColorableSubgraph}$, \cite{cook2019quantum} looks at $\mathtt{MaxkVertexCover}$ and \cite{niu2019optimizing} considers QAOA as a quantum state transfer protocol.

\section{QAMPA: Quantum Alternate ``Mixer-Phaser'' Ansatz}\label{sec:QAMPA}

We now introduce the ``Quantum Alternate Mixer-Phaser Ansatz'' (QAMPA) unitary, which is the ordered product of two-qubit gates:
\begin{eqnarray}
    \mathcal{\tilde U}(\gamma_p,\beta_p)&\equiv&
    \left(
    \prod_{n,m}\mathcal{U}_{MP}^{nm}(\gamma_p,\beta_p)
    \right) 
\end{eqnarray}
Here, $\mathcal{U}_{MP}^{nm}(\gamma_p,\beta_p)$ is a two-qubit operation between qubits $n$ and $m$ that is parameterized by two angles $\gamma_p, \beta_p \in \mathbb{R}$. The subscript $p$ refers to the $p$th round. We refer to this operation as a ``mixer-phaser'' (MP) operation in that it optionally implements mixing and phase separation depending on the value of the two independent parameters. A possible choice for the mixer-phase operation is simply\footnote{Another possibility could be $\mathcal{U}_{PS}^{nm}(\gamma_p)\mathcal{U}_{M}^{nm}(\beta_p)$.}: 
\begin{equation}
\mathcal{U}_{MP}^{nm}(\gamma_p,\beta_p)=\mathcal{U}_{M}^{nm}(\beta_p)\mathcal{U}_{PS}^{nm}(\gamma_p).    
\label{eq:MPproduct}
\end{equation}
We will focus on this choice for the analysis in this paper.

Note that
\begin{eqnarray}
    \mathcal{U}_{MP}^{nm}(\gamma_p,0)&=&\mathcal{U}_{PS}^{nm}(\gamma_p)\nonumber\\
    \mathcal{U}_{MP}^{nm}(0,\beta_p)&=&\mathcal{U}_{M}^{nm}(\beta_p).
    \label{eq:limits-QAMPA}
\end{eqnarray}
In other words, Eqs.~(\ref{eq:limits-QAMPA}) show that QAOA with $2p$ parameters $\gamma_{1},\dots, \gamma_{p}$, $\beta_{1},\dots, \beta_{p}$ can be mapped to QAMPA with $4p$ parameters if $\beta_{2k+1}=0$ and $\gamma_{2k}=0$ for $k=0\dots p$ (and the non-zero parameters are identified in sequence). However,  for the same number of layers the algorithm has double the parameters. It is not clear a priori how QAOA and QAMPA would perform relative to each other when compared at fixed, equal number of parameters. In the next subsection we will investigate this question numerically for a specific illustrative problem.

\subsection{Application to binary optimization with cardinality constraints}\label{subsec:QUBOconst}

Let's consider a special case of an integer program, a quadratic binary optimization problem  with $N$ variables, with a constraint that restricts the feasible subspace to bitstrings with a certain fixed Hamming weight $\kappa$. Mapping bits ({0, 1}) into spin variables ({-1, +1}), this results in the following Ising cost Hamiltonian and constraints:
\begin{eqnarray}
H_C&=&\sum_{nm}J_{nm}Z_n Z_m+\sum_n h_n Z_n\\
\text{subject to}&&\sum_n Z_n = 2(\kappa - N/2).
\label{eq:constraint}
\end{eqnarray}
For $h_n=0$, this problem is a weighted $\mathtt{MaxCut}$ with given sizes of partitions~\cite{ageev1999approximation} ($\mathtt{WeightedMaxCutGSP}$). It can also be seen as a Markowitzian portfolio optimization problem~\cite{markowitzportfolio} where the task is to select the best performing $\kappa$ of assets in a pool assuming correlations between their performance indicators. Note that while for small $\kappa$ the problem is clearly tractable, for the case $\kappa=N/2$ the problem turns into the NP-Hard $\mathtt{GraphBisection}$, which has been mapped to QAOA and studied numerically in the unweighted case where $J_{nm}$ are either 0 or 1 using a sparse XY-mixer in \cite{wilson2021optimizing}.

Following the literature, we construct the QAOA and QAMPA gates using $XY$ mixers~\cite{hadfield2019quantum, wang2019xy} as:
\begin{eqnarray}
\mathcal{U}_{PS}^{nm}(\gamma)=\mathcal{U}_{ZZ}^{nm}(\gamma)&=&e^{i\gamma\left(J_{nm}Z_n Z_m+h_nZ_n\right)}\\
\mathcal{U}_{M}^{nm}(\beta)=\mathcal{U}_{XY}^{nm}(\beta)&=&e^{i\beta\left(X_n X_m+Y_n Y_m\right)}\nonumber\\
\mathcal{\tilde{U}}_{MP}^{nm}(\gamma,\beta)&=&e^{i\left[\beta\left(X_n X_m+Y_n Y_m\right)+\gamma(J_{nm}Z_n Z_m+h_nZ_n)\right]}.\nonumber
\label{eq:gates}
\end{eqnarray}

The initial state could be taken to be an equal superposition of all solutions with $\kappa$ variables set to 1 on the qubit registers. i.e. a Dicke state~\cite{bartschi2019deterministic}.
By observing the periodicity of the unitaries composing the ans\"atze, we can observe that if the possible values of the coefficients are commensurable, the angle parameters could be selected within the domains $\gamma \in [0,2 \pi / \min_{>0} ( |h_n|, |J_{nm}|)]$ and $\beta \in [0,\pi]$ without loss of generality.

\subsection{Synthesis and routing}\label{subsec:synthesis}

Efficiently compiling a quantum circuit such as Eq.~\eqref{eq:ansatz} to a real quantum processors, having pre-defined calibrated two-qubit gates active on a sparse subset of all possible pairs of qubits, is a non-trivial planning and scheduling problem~\cite{venturelli2018compiling}. We would like to estimate the advantage of using QAMPA versus QAOA in common implementation scenarios. For $\mathtt{WeightedMaxCutGSP}$, the required $U_{ZZ}$ gates to implement the objective function are the $N (N - 1) / 2$ edges of a fully-connected graph. The mixer that is responsible for the exploration step of the algorithm by keeping the constraint~\eqref{eq:constraint} in check is also ideally a complete mixer, since it is proven numerically to be the best choice for Hamming weight constraints~\cite{wang2019xy,cook2019quantum,bartschi2020grover}. Choosing a mixer with sparser connectivity between various terms might lead to shorter circuits, but the compilation advantage of using QAMPA versus QAOA is maximal if we use the same graph for both phase-separation and mixing operations.

We should note that the initialization choice Eq.~\eqref{eq:init} (e.g. the creation of the Dicke state $|\psi_0\rangle$, which requires in principle $\mathcal{O}(\kappa N)$ gates~\cite{bartschi2019deterministic}), while being the simplest to analyze and possibly the most advantageous based in prior studies of similar problems~\cite{wang2019xy}, might be impractical in the near-term. As discussed in \cite{hadfield2019quantum, egger2020warm}, the initialization procedure  could be possibly substituted by a simpler to realize superposition of feasible states or a classical warm start candidate followed by a mixing round in QAOA. In QAMPA, the first round contains mixing so initialization might come for free if the gate is appropriately synthesized. Hence, initialization is not a concern for the discussion around compilation efficiency.

The routing requirements to schedule gates between all possible pairs of qubits depend on the underlying topology where swap operations can be performed. For a linear device, it was shown in \cite{kivlichan2018quantum} that the most efficient swap network allowing the scheduling of the gates could be executed with maximum parallelization in $N$ steps. For a more connected topology, the linear result is still a worst case scenario which can be implemented by defining an arbitrary Hamiltonian path on the device graph.

\begin{figure}
    \centering
    \includegraphics[width=0.95\columnwidth]{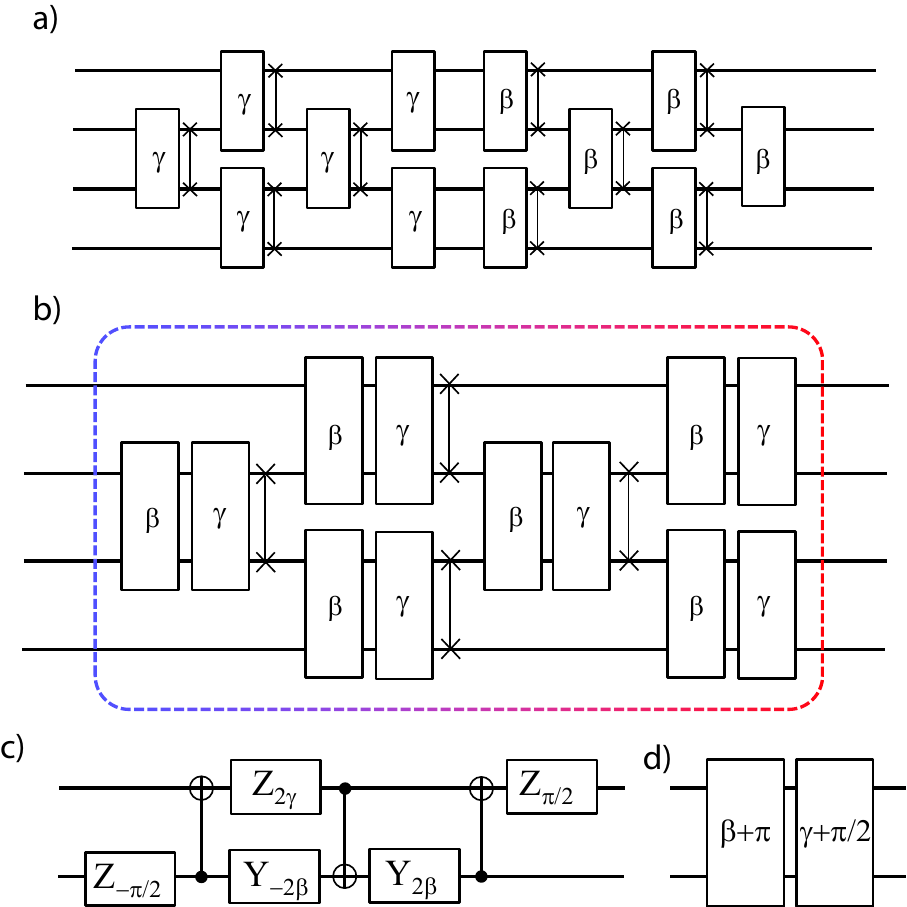} \\
    \caption{A pictorial view of the compiled circuits of (a) QAOA and (b) QAMPA for $p=1$ and a swap network that compiles all possible pairs of interactions efficiently on a line. c) shows the canonical decomposition of the mixer-phaser gate with additional swapping  $\mathtt{SWAP}\mathcal{U}_{MP}^{nm}(\gamma,\beta)$ into 3 CNOTs plus single qubit rotations. (d) the same gate decomposed using XY and ZZ as native gates. ($J_{nm}=1$ for simplicity in all figures)}
    \label{fig:circuits}
\end{figure}

As shown in Fig.~\ref{fig:circuits}-(a) for an illustratory $N = 4$ case, the linear efficient compilation of QAOA $p = 1$ is intertwined with a swap network both for the phase-separation layer (blue box) and for the mixing layer (red box). The routing overhead in this case is a total of $p (N - 1) ^ 2$ $\mathtt{SWAP}$ gates increasing the depth of about $2 p (N - 1)$ if the $\mathtt{SWAP}$ gates are not simplified or optimized in synthesis. For QAMPA instead, as shown in Fig.~\ref{fig:circuits}-(b), a single swap network is required for the mixing and phase separation layer, resulting in a clear advantage for circuit depth for the same number of parameters.

The optimal synthesis of logical gates depends on the available native operations on the quantum processor. For the sake of illustration, suppose to have access to the common set consisting of CNOT gates and parameterized single qubit rotations about $X$, $Y$ and $Z$. This set is universal and admits optimal synthesis formulas for any two-qubit gate utilizing at most 3 CNOTs and 15 single-qubit rotations~\cite{vatan2004optimal}.
Fig.~\ref{fig:circuits}-(c) illustrates the optimal synthesis of $\mathtt{SWAP}\mathcal{U}_{MP}^{nm}(\gamma,\beta)$ in terms of the canonical known decomposition.
Similar synthesis can be derived for $\mathtt{SWAP}\mathcal{U}_{ZZ}^{nm}$ and $\mathtt{SWAP}\mathcal{U}_{XY}^{nm}$, showing that for the pictured case there should be a factor of 2 between the resulting depths of the two ans\"atze. A different set of hardware primitives might increase or reduce the advantage, for instance if the $\textrm{fsim}$ gate~\cite{harrigan2021quantum} or the $XY$ gates are available natively then the swap could be subsumed in a renormalization of the angles, as illustrated in Fig.~\ref{fig:circuits}-(d) where we pictorialize the identity: $\mathtt{SWAP}\mathcal{U}_{MP}^{nm}(\gamma,\beta)=\exp(i\pi/4)\mathcal{U}_{ZZ}^{nm}(\gamma+\pi/2)\mathcal{U}_{XY}^{nm}(\beta+\pi)$.

If the underlying connectivity on the hardware is all-to-all, a swap network is not required. Still, for fixed number of angles it could still be depth-advantageous%
\footnote{We note that while depth correlates with fidelity of circuits, it has been suggested that real-time duration is a better metric to use to design compilers~\cite{venturelli2018compiling}.} %
to run QAMPA if the sum of the depth required for the synthesis of both the $\mathcal{U}_{PS}^{nm}$ and $\mathcal{U}_{M}^{nm}$ is larger than the depth required to synthesize $\mathcal{U}_{MP}^{nm}$, which is almost always the case if the QAOA gates are not natively available.

\section{Numerical evaluation}\label{sec:eval}

We benchmark the algorithms by numerically simulating the circuits for 40 random fully-connected $\mathtt{WeightedMaxCutGSP}$ problems where $J_{ij} \in \{-1,-0.5,0.5,1\}$ (chosen uniformly) and $h_i=0$, for even $N={4,6,\dots,16}$ and for $\kappa=N/2$ (representing the largest search space, $|\mathcal{F}|\simeq2^N/\sqrt{N}$). For simplicity, we set linear Zeeman terms $h_j = 0$, since it is mostly inconsequential from the perspective of the compilation overhead.
While the order of the execution of $\mathcal{U}_{ZZ}^{nm}$ gates does not matter, different orderings of the $\mathcal{U}_{XY}^{nm}$ and $\mathcal{U}_{MP}^{nm}$ are not equivalent. We consider that all runs related to a given instance are performed with a random permutation of the gates chosen among the sequences that are allowing maximum parallelization and minimum depth when intertwined with a swap network.

\subsection{Performance metric and parameter setting}\label{subsec:paramset}

\begin{figure*}[ht]
\begin{minipage}[b]{0.6\linewidth}
\centering
\includegraphics[width=\textwidth]{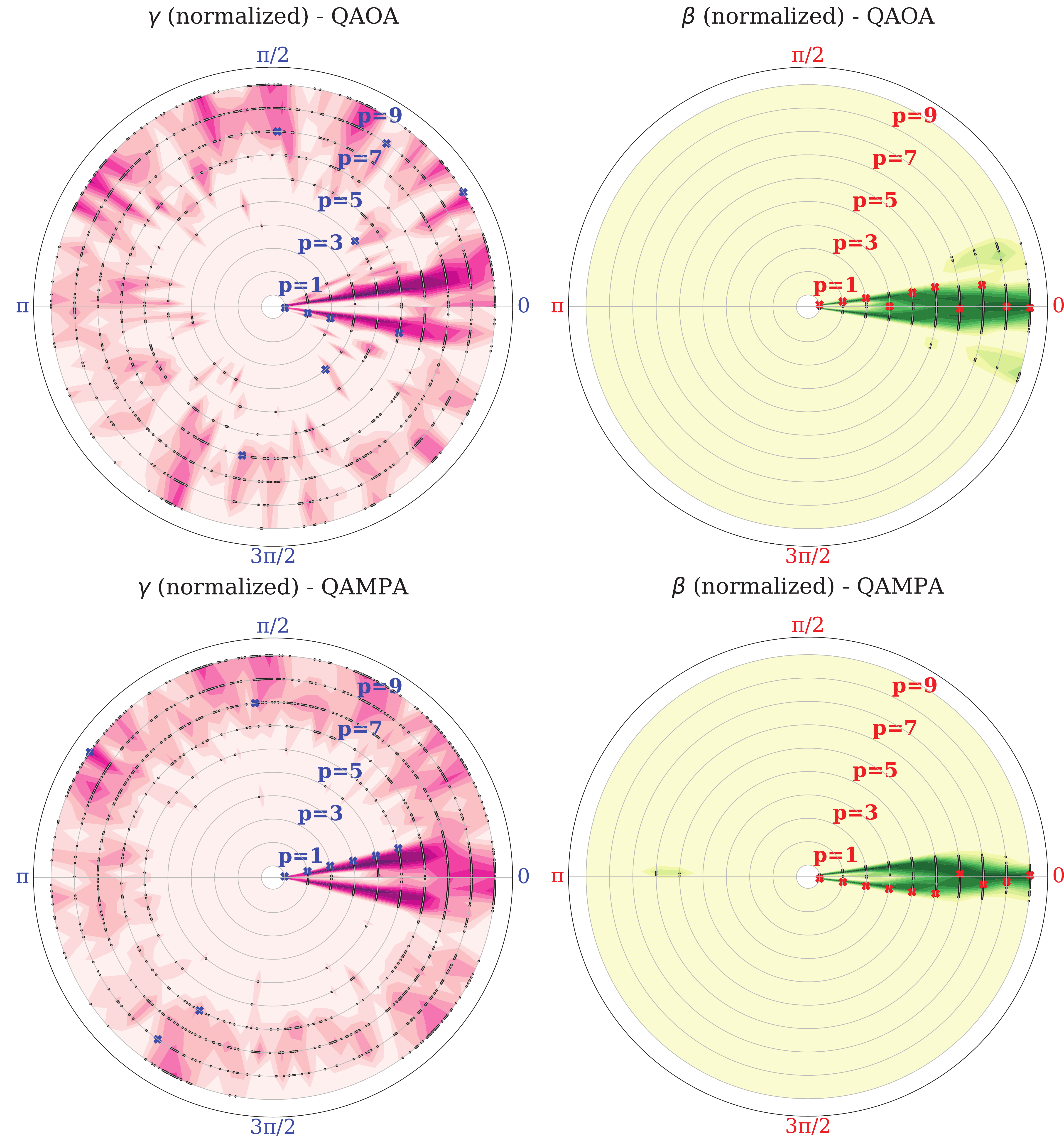}
\caption{Illustrative $\mathtt{scanlast}$ parameter setting procedure results for $\gamma$ (left column) and $\beta$ (right column) illustrated for an ensemble of 40 N=16 $\mathtt{WeightedMaxCutGSP}$ random instances for the procedure stopped at p=10. The colored heatmap reflects visually an interpolation of the probability density function for the value of top 10\% best performing angles (normalized in the range $0-2\pi$). The actual values for each instance at each iteration are marked as black dots. Blue (red) marks are the best found $\gamma$ ($\beta$) for the specific instance that returned the lowest metric score.}
\label{fig:polar-plots}
\end{minipage}
\hspace{0.5cm}
\begin{minipage}[b]{0.35\linewidth}
\centering
\includegraphics[width=\textwidth]{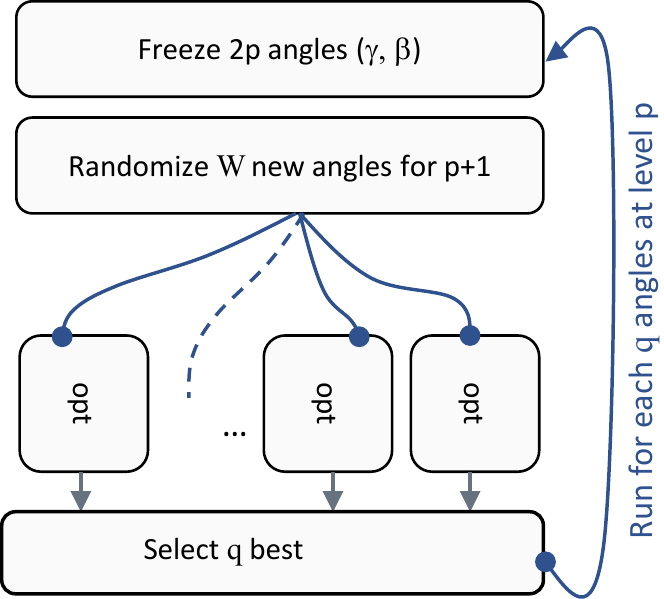}
\caption{The $\mathtt{scanlast}$ protocol. Note that at each iteration the two combinations of angles \{$\gamma_p$=0, $\beta_p$=0\} as well as \{$\gamma_p$=$\gamma_{p-1}$, $\beta_p$=$\beta_{p-1}$\} are included in each iteration.}
\label{fig:scanlast}
\includegraphics[width=\textwidth]{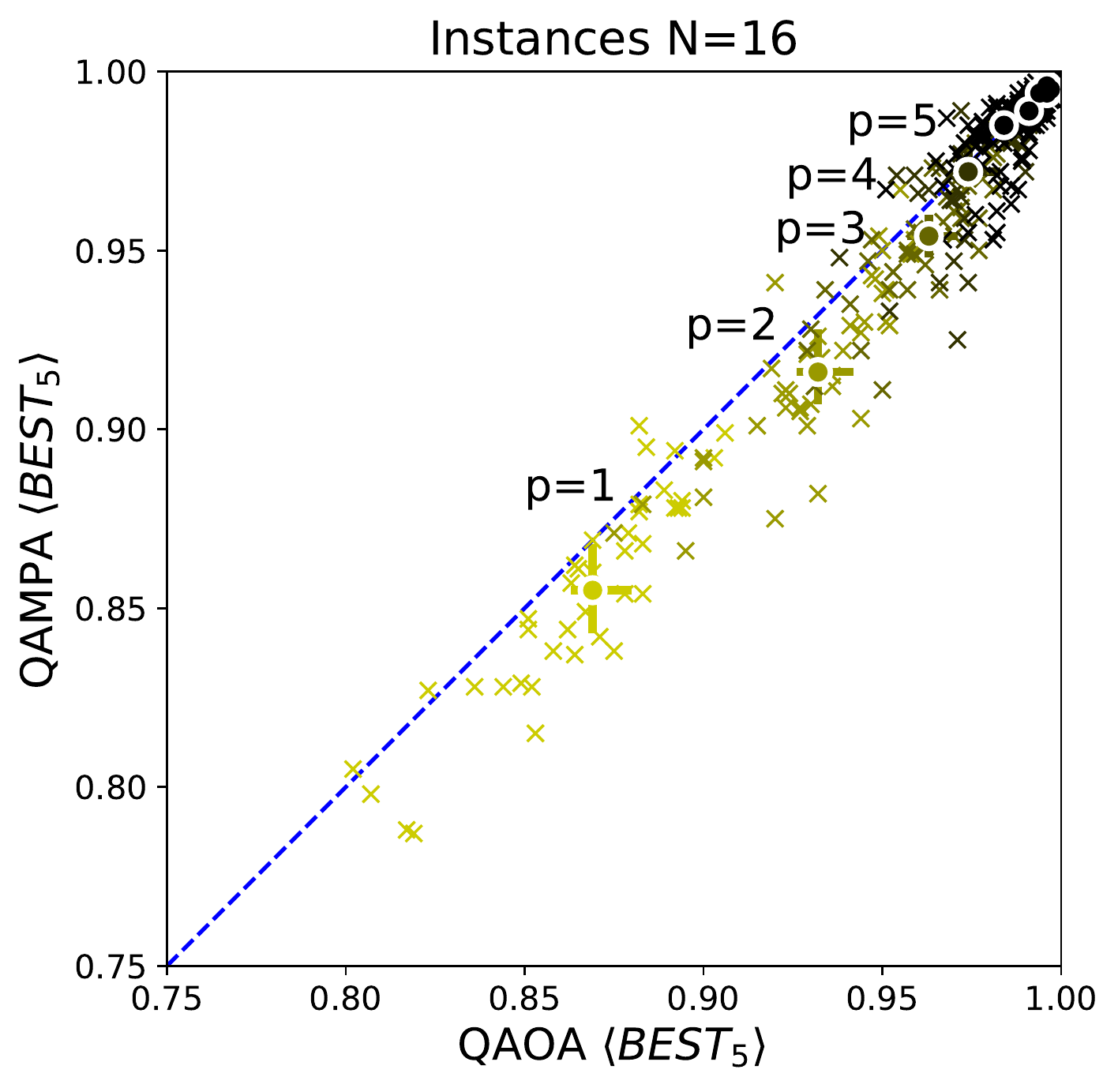}
\caption{Results obtained for the $\langle BEST_5 \rangle$ comparing instance-by-instance QAOA vs QAMPA. 
(bottom) N=16 as a function of p (darker marks indicate higher p=6-9).}
\label{fig:scatter-battle}
\end{minipage}
\end{figure*}

We want to evaluate the optimization performance of QAOA and QAMPA in terms of their value as a $\mathtt{WeightedMaxCutGSP}$ solver. The metric we use is related to a goal of running the algorithm  to discover as quickly as possible a bitstring associated to a sufficiently good value (as determined by a pre-defined quality threshold) of the objective function. To address this goal, our target performance metric will be the expected value of the objective function when the best result in $R$ runs is selected~\cite{kim2019leveraging}:
\begin{equation}\label{eq:BESTR}
     \langle \mathrm{BEST}_R\rangle = \sum_{k\in\mathcal{F}} \epsilon_k \left[\left(1- F(\epsilon_{k-1})\right)^R - \left(1-F(\epsilon_k)\right)^R \right],
\end{equation}
where $|k \rangle$ is a feasible state whose normalized objective function value is: 
\begin{equation}\label{eq:normalized_E}
\epsilon_k=\frac{\langle k| H_C |k \rangle-\epsilon_0}{\epsilon^\star-\epsilon_0}. 
\end{equation}
with $\epsilon_0$, $\epsilon^\star$ being respectively the minimum and the maximum of the objective function spectrum of values. $F(\epsilon_k)=\sum_{p<k}|\langle \psi_F(\vec\gamma,\vec\beta) | p \rangle|^2$ is the cumulative distribution function of $\epsilon_k$, where all sum over states are meant in ascending order $\epsilon_k$. It could be straightforwardly computed by accessing the wavefunction of the final state, or its sampling statistics.
This metrics, beside being the most relevant for optimization purposes, inherits the advantage discussed in the context of other metrics that are more focused on the high-quality solutions portion of the probability, such as the Conditional Value at Risk (CVaR)~\cite{barkoutsos2020improving, diez2021quantum} or Gibbs averages~\cite{li2020quantum}, which are suspected to have some desirable ``trainability'' properties to guide parameter setting, as opposed to the more traditional $\langle \psi_F(\vec\gamma,\vec\beta) | H_C | \psi_F(\vec\gamma,\vec\beta)\rangle$~\cite{mcclean2018barren} (which is simply $\langle\mathrm{BEST}_1\rangle$). For illustration, we will work with $\langle\mathrm{BEST}_5\rangle$, since $R=5$ seems to be a reasonable value to use to reach good approximation ratios for the moderate sizes of problems that we are studying, as we will demonstrate empirically ex-post.

The parameter setting strategy of choice for all experiments in this paper (which we call $\mathtt{scanlast}$, for easiness of reference - see Figure \ref{fig:scanlast}) follows a layerwise constructive optimization protocol employing an external blackbox optimizer that guides the repeated execution of the quantum circuit. The protocol aims to identify good angles for both QAOA and QAMPA at level $p+1$ using the information of the best found at level $p$. It starts with a random generation of $W_0$ pairs of angles that are then used as an initialization for each run of the optimizer.
The $q$ best found results ($\gamma_1^\star(q)$ and $\beta_1^\star(q)$) are then going to be each seeding the runs at $p=2$. More precisely, the $p=2$ runs will be each seeded by $q$ batches of $W$ runs each of the form $(\gamma_1^\star(q), \beta_1^\star(q), \gamma_2, \beta_2)$ where the last two angles are chosen randomly. Note that the best initial angles from the previous layer (e.g., $\gamma_1^*$ and $\beta_1^*$) are allowed to vary when optimizing the next layer.
The full layerwise procedure is applying this rule recursively: at layer $p+1$ we would launch $q$ searches where each run would initialize the optimizer with $(\gamma_1^\star(q), \beta_1^\star(q), \dots, \gamma_p^\star(q), \beta_p^\star(q), \gamma_{p+1}, \beta_{p+1})$ for a total complexity of $\mathcal{O}((W_0 + pqW)f_{opt})$ where $f_{opt}$ is the number of function evaluations used per optimization attempt.

For most of our tests, we choose $W_0 = 50$, $q=10$, $W=250$ and $f_{opt} = 250$, for a total number of runs of $625000+2500p$ per test. Moreover we decide to use Powell's method for the external optimization loop. While this method is not expected to be a suitable choice for large number of variables, in our problem set it outperformed several other methods used in the literature (e.g. BFGS, Model Gradient Descent, and Nelder-Mead) and provided the closest results to bruteforce search. In Fig.~\ref{fig:polar-plots} we illustrate the results of the $\mathtt{scanlast}$ procedure in practice, showing the 10\% best-found QAOA/QAMPA for each instance (black dots) that maximize $\langle \mathrm{BEST}_5 \rangle$ for 40 instances with $N=16$ variables.  
The displayed results illustrate clearly that for a given instance the best-found phase separation $\gamma_p^\star$ and mixing angles $\beta_p^\star$ for QAOA and QAMPA are similar, at least for low value of $p$.  

Indeed, by analyzing data from random instances of sizes we observe that the first angles of the sequence converge very often to a value of constant magnitude, which is often close to 0.  The time-reversal symmetry is manifest in the heatmap since the solutions for which all the angles are negated are equivalent, and the choice of one over another is likely associated to the randomness in $\mathtt{scanlast}$. For the mixing angle the concentration of best performing angles is apparent for almost all tested instances, while for the $\gamma$ angles 
at $p>6$ the concentration is not as striking. This could be explained by the fact that the value of the $\langle BEST_5 \rangle$ metric is already very close to the maximum, so the optimization landscape could be rather flat and difficult to numerically optimize.

\subsection{Instance-by-instance comparison}\label{subsec:comparison}

To compare the performance of the two ansatz we consider the instance-by-instance scatter plot where we compare the $\langle \mathrm{BEST}_5 \rangle$ results for QAOA vs QAMPA after the $\mathtt{scanlast}$ optimization.\footnote{Note that we do not consider the runs that are required to discover the best performing angles but we directly report the metric computed on the corresponding $F(\epsilon_k)$ distribution.} The figure indicates with cross symbols the metric value for each instance and with the round dots the median/standard deviation of the results for all the instance results associated to a given $p$. What is clear from the results shown in Figure~\ref{fig:scatter-battle} is that as $p$ increases both ans\"atze perform essentially the same for all tested sizes. This convergence is not surprising as for large $p$ QAOA and QAMPA for the same number of parameters could be interpreted as a Trotter-like approximation of the same unitary evolution~\cite{childs2021theory}. Similar results are observed also for different $R$ in the metric (including the common expectation value metric $R=1$), indicating that the output distribution for the $\epsilon_k$  at the current sizes is smooth and monotonic.

\begin{figure*}[ht]
\includegraphics[width=\textwidth]{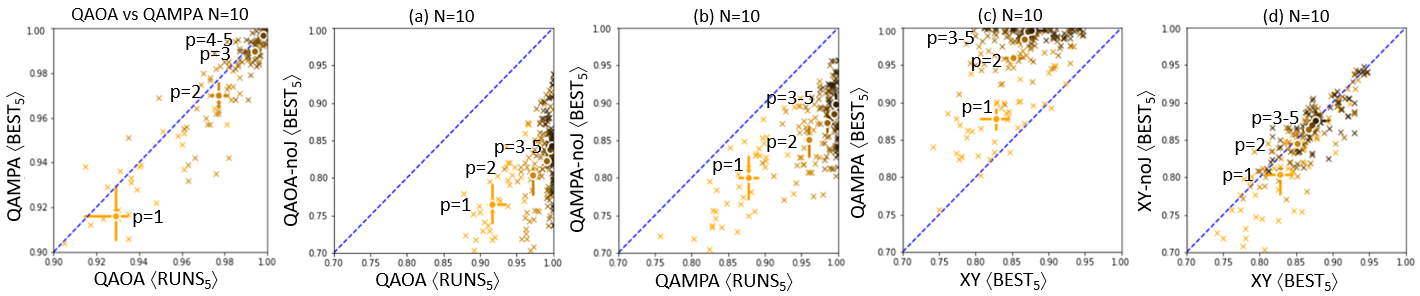}
\caption{N=10 QAOA vs QAMPA instance-by instance comparison (left) and Ansatz Variations (a)-(d). All results are after the $\mathtt{scanlast}$ parameter setting. QAOA-noJ in (a) is constructed with the phase-separation gates $\mathcal{U}_{PS}^{nm}(\gamma,\beta)|_{J_{nm}=1}$, the QAMPA-noJ ansatz in (b) is constructed with the mixer-phaser gates $\mathcal{\tilde{U}}_{MP}^{nm}(\gamma,\beta)|_{J_{nm}=1}$, and the XY-noJ ansatz in (d) uses $\mathcal{\tilde{U}}_{XY}^{nm}(\gamma,\beta)|_{J_{nm}=1}$. In (c) it is shown that the simplified circuit with $\gamma_p=0$ (XY) is not performing well against other options.}
\label{fig:temp-extra}
\end{figure*}

\subsubsection*{Additional tests and variations of QAMPA/QAOA}

The approach that we benchmarked in this study offers some flexibility on its implementation. For instance, given that QAMPA is not fully grounded on insights from a specific Hamiltonian evolution, it might be worth asking whether the information related to the coefficient of objective function ($J_{nm}$) are beneficial as inserted in the circuit or if it is sufficient to train the parameters using that information, like done for VQE hardware-efficient ans\"atze~\cite{kandala2017hardware}.
We investigated empirically these questions by comparing instance-by-instance results for QAOA and QAMPA, using $\mathtt{scanlast}$ on a set of instances for $N=10$ with slightly modified parameters ($f_{opt}=200p$). The results are statistically in line with the ones presented for $N=16$ in Figure \ref{fig:scatter-battle} using higher computational effort. 

In Figure \ref{fig:temp-extra}-(a),(b) we show results for algorithms against a version where for the circuit ansatz all $J_{nm}$ are fixed to be 1 (we call this the -noJ version in the figures).\footnote{Similar results are obtained if instead of putting all coefficients to be equal we set them to be proportional to a random number drawn from the same distribution that generated the instance in the first place.} Note that the cost function coefficients are still used in the evaluation metric both for the parameter setting and for scoring the performance. Another test we performed, illustrated in Figure \ref{fig:temp-extra}-(c),(d), compares the QAMPA performance against the performance of an ansatz that includes only $XY$ gates. For the \emph{XY-noJ} variation case, the phase separation step of QAOA is completely eliminated and the phase information is entirely absent. The $XY$ label in the figure considers instead the design of an XY-only ansatz that mixes qubits using an angle proportional to the cost function coefficients, i.e. using gates of the form $\mathcal{U}^{nm}_{XY}(J_{nm} \beta)$.

What is observed is that the standard QAOA approach (which performs only slightly better than QAMPA, as we recall) gives the best performance compared against all other tested variations, and it is always beneficial to include a proportional factor multiplying $\gamma_p$, for each gate between qubits $n$ and $m$, corresponding to the objective function coefficients $J_{nm}$ in the circuit ansatz. These observations were validated for multiple problem sizes up to $N=16$.

\section{Discussion and Conclusions}

As reviewed in \cite{cerezo2020variational} and \cite{bharti2021noisy}, modern quantum algorithms for optimization on NISQ devices have generalized significantly the original structure of the QAOA circuitry.
In this paper, we have presented a variation that combines the hardware-efficient spirit of Variational Quantum Eigensolvers with the advanced mixers of the Quantum Alternating Operator Ansatz and the guidance from inclusion of operators derived from the cost function without increasing the number of parameters that need to be optimized.   
Our numerical results indicate that the mixer-phaser ansatz QAMPA is a compelling choice among NISQ era quantum options;
we expect these ideas can be ported beyond the
 $\mathtt{WeightedMaxCutGSP}$ problem, to provide compilation advantages to a multitude of other hard-constrained combinatorial optimization problems that require advanced mixers. Multiple questions however remain in order to bridge the gap from proof-of-principles to real-world implementation. We list here few avenues of research towards the goal of deploying a QAMPA solver in a quantum processor, before concluding with some more general thoughts on research directions.

First, the parameter setting procedure used in this study could be refined to avoid optimization bottlenecks and limitations that affect all layerwise training protocols at scale. In particular, the Powell method while effective at small $N$ would eventually become intractable, and alternative gradient methods might be hampered by barren plateaus if the parameter setting protocol is kept to be ``layerwise''~\cite{campos2020abrupt}. Recently developed analytical methods based on series expansion ~\cite{hadfield2021analytical} or quantum control ~\cite{larocca2021diagnosing} methods might come in handy to analyze further the reachability deficits and strengths of these algorithms~\cite{akshay2020reachability}, study the observed optimal parameter concentration~\cite{akshay2021parameter}, and to estimate the performance of QAMPA at scale.
Ties between quantum annealing schedule and QAOA parameter setting \cite{yang2017optimizing,zhou2020quantum,brady2021optimal},
further indicating that cross-overs between digital and analog optimization methods are also an interesting possible development for QAMPA~\cite{magann2021pulses, headley2020approximating, wiersema2020exploring, brady2021behavior}.

Moreover, our performance evaluation procedure based on the $\langle \mathrm{BEST}_5 \rangle$ provides just an indication of the ability of a QAMPA solver to identify a good solution of $\mathtt{WeightedMaxCutGSP}$ in a reasonable time, and a fully-fledged analysis on comparative advantage of using this method versus other heuristics is required. This analysis has to take into account practical issues such as the effect of noise, which is going to adversely impact our performance estimation. The relative performance results will be both problem and hardware dependent. While theoretical frameworks to estimate the impact of noise in circuits featuring XY gates are being developed~\cite{streif2021quantum}, ultimately only the experimental tests on quantum hardware will be able to provide a full picture of performance, including identifying at what layer $p$ the solver is most effective. While in the ideal case, performance cannot decrease with $p$, under noise the situation is different; performance is observed to decrease quickly with $p$ in current hardware~\cite{harrigan2021quantum}. 
QAMPA, with its circuit depth reduction compared to QAOA,  will enable empirical studies for higher values of $p$ than QAOA on a wide variety of quantum hardware.
Hybridization of QAMPA with adaptive~\cite{zhu2020adaptive} and recursive~\cite{bravyi2020obstacles} hybrid methods may prove powerful. One advantage of the studied problem, and of others with locally conserved particle number constraints, is that provides a natural error mitigation strategy via post-selection~\cite{shaydulin2021error}. Namely, measured bitstrings which do not obey the Hamming weight constraint are discarded. Post-selection has shown to provide significant improvements in experiments on superconducting processors~\cite{google2020hartree} and can be generally applied beneficially to any situation with constraints. Another recent experiment~\cite{hashim2021optimized} shows that permuting the ordering of the qubits in the SWAP network and averaging over the results is reducing the systematic coherent errors, a technique that could be generalized to our case where the permutations are not equivalents, possibly helping the parameter setting and/or optimization performance.

In terms of actual implementation, the QAMPA method is readily testable in superconducting processors that natively support XY interactions, such as Rigetti's devices of the Aspen family~\cite{abrams2019implementation}. However, the initialization to a Dicke state is likely too heavy for near-term implementation, so a warm-start from a classical candidate solution obtained by greedy search is advisable~\cite{egger2020warm}. 

We expect researchers to develop other ans\"atze that have sweet spots in terms of circuit depth, number of parameters, and performance. As QAMPA illustrates, there is nothing sacrosanct about QAOA; variants may perform as well or better, particularly on NISQ hardware. A rich ecosystem of ans\"atze that take into account hardware architectures, gate sets, and noise considerations for different types of NISQ processors will enable more rapid understanding of quantum optimization approaches for the NISQ era and beyond.

\begin{acknowledgments}

R.L. acknowledges support from a NASA Space Technology Graduate Research (NSTGRO) Fellowship, AFRL NYSTEC Contract (FA8750-19-3-6101), and the USRA Feynman Quantum Academy, a program of USRA NASA Academic Mission Services (NNA16BD14C). D.V. also acknowledges support of NAMS. All authors acknowledge support from the DARPA ONISQ program (IAA 8839 and agreement No. HR00112090). All authors acknowledge useful discussion with the NASA Quantum AI Laboratory (QuAIL) team members, in particular Stuart Hadfield, and Benjamin Hall from Michigan State University.

\end{acknowledgments}

\bibliography{biblio}

\end{document}